# RANDOM SURVIVAL FORESTS[1]

By Hemant Ishwaran, Udaya B. Kogalur,
Eugene H. Blackstone and Michael S. Lauer

*Cleveland Clinic, Columbia University, Cleveland Clinic and National Heart, Lung, and Blood Institute*

We introduce random survival forests, a random forests method for the analysis of right-censored survival data. New survival splitting rules for growing survival trees are introduced, as is a new missing data algorithm for imputing missing data. A conservation-of-events principle for survival forests is introduced and used to define ensemble mortality, a simple interpretable measure of mortality that can be used as a predicted outcome. Several illustrative examples are given, including a case study of the prognostic implications of body mass for individuals with coronary artery disease. Computations for all examples were implemented using the freely available R-software package, `randomSurvivalForest`.

**1. Introduction.** In this article we introduce *random survival forests*, an ensemble tree method for analysis of right-censored survival data. As is well known, constructing ensembles from base learners, such as trees, can substantially improve prediction performance. Recently it has been shown by Breiman (2001) that ensemble learning can be improved further by injecting randomization into the base learning process, an approach called *random forests*. Random survival forests (RSF) methodology extends Breiman's random forests (RF) method. In RF, randomization is introduced in two forms. First, a randomly drawn bootstrap sample of the data is used to grow a tree. Second, at each node of the tree, a randomly selected subset of variables (covariates) is chosen as candidate variables for splitting. Averaging over trees, in combination with the randomization used in growing a tree, enables RF to approximate rich classes of functions while maintaining low generalization error. Considerable empirical evidence has shown RF to be

Received January 2008; revised March 2008.
[1]Supported in part by National Institutes of Health RO1 Grant HL-072771.
*Key words and phrases.* Conservation of events, cumulative hazard function, ensemble, out-of-bag, prediction error, survival tree.

highly accurate, comparable to state-of-the-art methods such as bagging [Breiman (1996)], boosting [Schapire et al. (1998)], and support vector machines [Cortes and Vapnik (1995)].

Until now, applications of RF have focused primarily on classification and regression problems. Even the popular R-software package randomForest [Liaw and Wiener (2002, 2007)] considers only regression and multiclass data settings, not survival analysis. Extending random forests to right-censored survival data is of great value. Survival data are commonly analyzed using methods that rely on restrictive assumptions such as proportional hazards. Further, because these methods are often parametric, nonlinear effects of variables must be modeled by transformations or expanding the design matrix to include specialized basis functions. Often ad hoc approaches, such as stepwise regression, are used to determine if nonlinear effects exist. Identifying interactions, especially those involving multiple variables, is also problematic. This must be done by brute force (examining all two-way and three-way interactions, e.g.), or must rely on subjective knowledge to narrow the search.

In contrast, these difficulties are handled automatically using forests. We illustrate the ease with which RSF can uncover complex data structures through an in-depth case study of the prognostic implications of being underweight, overweight, or obese and having severe, but stable coronary artery disease. Although much has been written about public health ramifications of the obesity epidemic [Olshansky et al. (2005)], considerable controversy exists regarding the precise association of body mass with prognosis. Investigators have noted complex patterns surrounding possible reverse causation in underweight individuals, interactions with smoking, and an unclear inflection point at which point increasing body mass confers increased risk [Adams et al. (2006), Flegal (2005, 2007), Fontaine et al. (2003)]. Some have identified a possible obesity paradox among patients with established heart disease in which increased body mass predicts better survival [Urtesky et al. (2007)]. To clarify these issues, we analyzed a large cohort of patients with coronary artery disease undergoing isolated coronary artery bypass surgery. Using RSF, we identified a complex relationship between long-term survival, body mass, renal (kidney) function, smoking, and number of internal coronary artery byass grafts. We believe our novel findings help explain some of the apparent contradictions previously reported.

1.1. *Other forest approaches.* RSF strictly adheres to the prescription laid out by Breiman (2003) and in this way differs from other forest approaches to survival data. Breiman's prescription requires that all aspects of growing a random forest take into account the outcome. In right-censored survival settings, this comprises survival time and censoring status. Thus, the splitting criterion used in growing a tree must explicity involve survival time and censoring information. Tree node impurity, measuring effectiveness of a split in separating data, must measure separation by survival difference. Further, the predicted value for a terminal node in a tree, the resulting ensemble predicted value from the forest, and the measure of prediction accuracy must all properly incorporate survival information.

This differs from other forest approaches to survival analysis that tend to be "off-the-shelf." These do not strictly implement Breiman's forest method [Breiman (2003)], but instead recast the survival setting into one that can be treated using existing forest methodology. For example, survival analysis is possible by working within the Classification and Regression Tree (CART) paradigm [Breiman et al. (1984)] by reformulating a survival tree in terms of a classification tree. One such example is Ishwaran et al. (2004), where under a proportional hazards assumption, CART methodology is used to produce a relative risk forest by exploiting an equivalence to Poisson tree likelihoods [LeBlanc and Crowley (1992)]. Another interesting approach, considered by Hothorn et al. (2006), analyzes right-censored survival data by using log-transformed survival time as the outcome in a weighted RF regression analysis. Observations in the regression analysis are weighted by, what are referred to as, inverse probability of censoring (IPC) weights. See also Molinaro, Dudoit and van der Laan (2004).

Early experimental work by Breiman (2002) on survival forests is also relevant. In this approach a survival tree is grown using a hybrid splitting method in which nodes are split both on time and covariates. This yields a nonparametric estimate for the survival function that can then be used to trace the effects of variables on survival as a function of time.

1.2. *Objectives and outline.* The purpose of this article is to give a detailed description of RSF (Sections 2 and 3) and to illustrate several of its important features. A core idea underlying the approach is a conservation-of-events principle for survival trees (Section 4). This principle is used to define ensemble mortality, a new type of predicted outcome for survival data. Ensemble mortality has a natural interpretation in terms of the expected total number of deaths and is derived from the ensemble cumulative hazard function (CHF), the forest predicted value for the CHF. Prediction error is defined in Section 5, and prediction accuracy of the ensemble is compared with several competing methods in a large experiment comprising several real and simulated datasets (Section 6). Our results are quite promising and add to the growing list of successful applications of RF. In Section 7 we investigate the use of variable importance for variable selection. In Section 8 we introduce a novel missing data algorithm for forests. The algorithm can be used for both training and test data and applies to missing values for both covariates and survival data. The paper ends with our case study of body mass of patients with coronary artery disease (Section 9).

**2. Random survival forests algorithm.** We begin with a high-level description of the algorithm. Specific details follow:

1. Draw $B$ bootstrap samples from the original data. Note that each bootstrap sample excludes on average 37% of the data, called out-of-bag data (OOB data).
2. Grow a survival tree for each bootstrap sample. At each node of the tree, randomly select $p$ candidate variables. The node is split using the candidate variable that maximizes survival difference between daughter nodes.
3. Grow the tree to full size under the constraint that a terminal node should have no less than $d_0 > 0$ unique deaths.
4. Calculate a CHF for each tree. Average to obtain the ensemble CHF.
5. Using OOB data, calculate prediction error for the ensemble CHF.

**3. Ensemble cumulative hazard.** Central elements of the RSF algorithm are growing a survival tree and constructing the ensemble CHF. Here we provide details necessary to understand these.

3.1. *Binary survival tree.* Similar to CART, survival trees are binary trees grown by recursive splitting of tree nodes. A tree is grown starting at the root node, which is the top of the tree comprising all the data. Using a predetermined survival criterion, the root node is split into two daughter nodes: a left and right daughter node. In turn, each daughter node is split with each split giving rise to left and right daughters. The process is repeated in a recursive fashion for each subsequent node.

A good split for a node maximizes survival difference between daughters. The best split for a node is found by searching over all possible $x$ variables and split values $c$, and choosing that $x^*$ and $c^*$ that maximizes survival difference. By maximizing survival difference, the tree pushes dissimilar cases apart. Eventually, as the number of nodes increase, and dissimilar cases become separated, each node in the tree becomes homogeneous and is populated by cases with similar survival.

3.2. *Terminal node prediction.* Eventually the survival tree reaches a saturation point when no new daughters can be formed because of the criterion that each node must contain a minimum of $d_0 > 0$ unique deaths. The most extreme nodes in a saturated tree are called terminal nodes. Denote these by $\mathscr{T}$. Let $(T_{1,h}, \delta_{1,h}), \ldots, (T_{n(h),h}, \delta_{n(h),h})$ be the survival times and the 0–1 censoring information for individuals (cases) in a terminal node $h \in \mathscr{T}$. An individual $i$ is said to be right-censored at time $T_{i,h}$ if $\delta_{i,h} = 0$; otherwise, if $\delta_{i,h} = 1$, the individual is said to have died (experienced an event) at $T_{i,h}$. Let $t_{1,h} < t_{2,h} < \cdots < t_{N(h),h}$ be the $N(h)$ distinct event times. Define $d_{l,h}$ and $Y_{l,h}$ to be the number of deaths and individuals at risk at time $t_{l,h}$. The CHF estimate for $h$ is the Nelson–Aalen estimator

$$\hat{H}_h(t) = \sum_{t_{l,h} \leq t} \frac{d_{l,h}}{Y_{l,h}}.$$

All cases within $h$ have the same CHF.

Each case $i$ has a $d$-dimensional covariate $\mathbf{x}_i$. The notation $x$ above refers to one coordinate of $\mathbf{x}_i$. Let $H(t|\mathbf{x}_i)$ be the CHF for $i$. To determine this value, drop $\mathbf{x}_i$ down the tree. Because of the binary nature of a survival tree, $\mathbf{x}_i$ will fall into a unique terminal node $h \in \mathscr{T}$. The CHF for $i$ is the Nelson–Aalen estimator for $\mathbf{x}_i$'s terminal node:

$$(3.1) \qquad H(t|\mathbf{x}_i) = \hat{H}_h(t), \qquad \text{if } \mathbf{x}_i \in h.$$

Identity (3.1) defines the CHF for all cases and defines the CHF for the tree.

3.3. *The bootstrap and OOB ensemble CHF.* The CHF (3.1) is derived from a single tree. To compute an ensemble CHF, we average over $B$ survival trees. We describe both an OOB and bootstrap estimate.

Recall that each tree in the forest is grown using an independent bootstrap sample. Define $I_{i,b} = 1$ if $i$ is an OOB case for $b$; otherwise, set $I_{i,b} = 0$. Let $H_b^*(t|\mathbf{x})$ denote the CHF (3.1) for a tree grown from the $b$th bootstrap sample. The OOB ensemble CHF for $i$ is

$$(3.2) \qquad H_e^{**}(t|\mathbf{x}_i) = \frac{\sum_{b=1}^B I_{i,b} H_b^*(t|\mathbf{x}_i)}{\sum_{b=1}^B I_{i,b}}.$$

Observe that (3.2) is an average over bootstrap samples in which $i$ is OOB. Equivalently, $H_e^{**}(t|\mathbf{x}_i)$ can be calculated as follows. Drop OOB data down a survival tree grown from in-bag (bootstrap) data. Keep track of $i$'s terminal node and its CHF. Take the average of these CHFs. This yields (3.2).

In contrast to (3.2), the bootstrap ensemble CHF for $i$ is

$$(3.3) \qquad H_e^*(t|\mathbf{x}_i) = \frac{1}{B} \sum_{b=1}^B H_b^*(t|\mathbf{x}_i).$$

Observe that (3.3) uses all survival trees and not just those where $i$ is OOB.

**4. Conservation of events.** We use (3.2) and (3.3) to define a predicted outcome. Our approach rests on a conservation-of-events principle [Naftel, Blackstone and Turner (1985)]. Under fairly general conditions, conservation of events asserts that the sum of the estimated CHF over observed time (both censored and uncensored) equals the total number of deaths. This applies to a wide collection of estimators, including the Nelson–Aalen estimator. For a terminal node $h \in \mathscr{T}$ in a given tree, conservation of events can be stated as the following lemma:

LEMMA 1.  $\sum_{i=1}^{n(h)} \hat{H}_h(T_{i,h}) = \sum_{i=1}^{n(h)} \delta_{i,h}$ *for each terminal node* $h \in \mathcal{T}$. *In other words, the total number of deaths is conserved within* $h$.

Lemma 1 shows that summing the CHF over observed survival times equals the total number of deaths within a terminal node: a type of conservation of events within the ends of a tree. An immediate corollary is the stronger assertion that the total number of deaths in a tree is also conserved. Let $(T_1, \delta_1), \ldots, (T_n, \delta_n)$ denote survival times and censoring values for the nonboostrapped data.

COROLLARY 1.   *For each tree grown from the original nonbootstrapped data,*

$$\sum_{i=1}^{n} H(T_i|\mathbf{x}_i) = \sum_{h \in \mathcal{T}} \sum_{i=1}^{n(h)} \hat{H}_h(T_{i,h}) = \sum_{h \in \mathcal{T}} \sum_{i=1}^{n(h)} \delta_{i,h} = \sum_{i=1}^{n} \delta_i.$$

*Note that the right-hand side equals total number of deaths.*

4.1. *Ensemble mortality.* Corollary 1 is used to motivate a predicted outcome measuring mortality. Corollary 1 shows that the total number of deaths equals the sum of the CHF over $(T_i, \mathbf{x}_i)$. Mortality, in contrast, is defined as the expected value for the CHF summed over time $T_j$, conditioned on a specific $\mathbf{x}_i$. It measures the number of deaths expected under a null hypothesis of similar survival behavior. Specifically, the mortality for $i$ is defined as

$$M_i = \mathbb{E}_i \left( \sum_{j=1}^{n} H(T_j|\mathbf{x}_i) \right),$$

where $\mathbb{E}_i$ is the expectation under the null hypothesis that all $j$ are similar to $i$.

Mortality can be estimated naturally within a survival tree paradigm. The structure of a survival tree enforces a null hypothesis of similar survival within its terminal nodes; individuals in a terminal node share a common estimated hazard function. The nature of a survival tree and its forest therefore suggests an estimate for mortality. We refer to this estimate as ensemble mortality. The ensemble mortality for $i$ is defined as

$$\hat{M}^*_{e,i} = \sum_{j=1}^{n} H^*_e(T_j|\mathbf{x}_i).$$

Similarly, OOB ensemble mortality is defined as $\hat{M}^{**}_{e,i} = \sum_{j=1}^{n} H^{**}_e(T_j|\mathbf{x}_i)$.

**5. Prediction error.** To estimate prediction error, we use Harrell's concordance index [Harrell et al. (1982)]. The C-index (concordance index) is related to the area under the ROC curve [Heagerty and Zheng (2005)]. It estimates the probability that, in a randomly selected pair of cases, the case that fails first had a worst predicted outcome. The interpretation of the C-index as a misclassification probability is attractive, and is one reason we use it for prediction error. Another attractive feature is that, unlike other measures of survival performance, the C-index does not depend on a single fixed time for evaluation. The C-index also specifically accounts for censoring.

5.1. *C-index calculation.* The C-index is calculated using the following steps:

1. Form all possible pairs of cases over the data.
2. Omit those pairs whose shorter survival time is censored. Omit pairs $i$ and $j$ if $T_i = T_j$ unless at least one is a death. Let Permissible denote the total number of permissible pairs.
3. For each permissible pair where $T_i \neq T_j$, count 1 if the shorter survival time has worse predicted outcome; count 0.5 if predicted outcomes are tied. For each permissible pair, where $T_i = T_j$ and both are deaths, count 1 if predicted outcomes are tied; otherwise, count 0.5. For each permissible pair where $T_i = T_j$, but not both are deaths, count 1 if the death has worse predicted outcome; otherwise, count 0.5. Let Concordance denote the sum over all permissible pairs.
4. The C-index, $C$, is defined by $C = \text{Concordance}/\text{Permissible}$.

5.2. *OOB prediction error.* Calculating $C$ requires a predicted outcome. We use the OOB ensemble CHF to define a predicted outcome similar to ensemble mortality described in Section 4.1. Because this value is derived from OOB data, it can be used to obtain an OOB estimate for $C$, and, consequently, an OOB error rate.

Let $t_1^o, \ldots, t_m^o$ denote pre-chosen unique time points (we use the unique event times, $t_1, \ldots, t_N$). To rank two cases $i$ and $j$, we say $i$ has *a worse predicted outcome* than $j$ if

$$\sum_{l=1}^{m} H_e^{**}(t_l^o|\mathbf{x}_i) > \sum_{l=1}^{m} H_e^{**}(t_l^o|\mathbf{x}_j).$$

Using this rule, compute $C$ as outlined above. Denote the OOB estimate by $C^{**}$. The OOB prediction error, $\text{PE}^{**}$, is defined as $1 - C^{**}$. Note that $0 \leq \text{PE}^{**} \leq 1$ and that a value $\text{PE}^{**} = 0.5$ indicates prediction no better than random guessing.

**6. Empirical comparisons.** Here we report the results of an experiment designed to study prediction accuracy of RSF. Prediction performance was calculated using the C-index, $C$, of Section 5.

Eleven datasets were used in the experiment; eight were distinct. One of these, node-positive breast cancer data studied in Hothorn et al. (2006), was used to create three additional datasets. To study stability in high-dimensional settings, 10, 50 and 100 uncorrelated variables representing noise were drawn from a uniform distribution and were added to the data. We refer to these data as breast10, breast50 and breast100, and the original dataset as breast. Of the remaining seven datasets, four are provided in the randomSurvivalForest R-package. These are as follows: veteran's administration lung cancer data from Kalbfleisch and Prentice (1980) (veteran); primary biliary cirrhosis data from Fleming and Harrington (1991) (pbc); burn patient data from Kalbfleisch and Prentice (1980) (burn); and recidivism data from Rossi, Berk and Lenihan (1980) (recid). The remaining three datasets were a prostate dataset described in Kattan (2003) (prostate), a dataset comprising patients listed for heart transplant at Cleveland Clinic (transplant), and early stage esophageal cancer data considered in Ishwaran et al. (2004) (esophagus).

Computations were implemented using randomSurvivalForest software under its default settings [Ishwaran and Kogalur (2007, 2008)]. In each instance 1000 trees were grown. Each of the four splitting rules available in the software package were used. These were as follows [for more details see Ishwaran and Kogalur (2008)]:

1. A log-rank splitting rule (logrank) that splits nodes by maximization of the log-rank test statistic [Segal (1988), LeBlanc and Crowley (1993)].
2. A conservation-of-events splitting rule (conserve) that splits nodes by finding daughters closest to the conservation-of-events principle.
3. A log-rank score rule (logrankscore) that splits nodes using a standardized log-rank statistic [Hothorn and Lausen (2003)].
4. A random log-rank splitting rule (logrankrandom). A random split is selected for each of the $p$ candidate variables in a node, and the variable with maximum log-rank statistic (at its random split point) is used to split the node.

As a benchmark, we used Cox regression. The RF approach of Hothorn et al. (2006) was also included for comparison. In this latter method, 1000 regression trees were grown. Each tree was derived from bootstrap data where each observation was sampled with probability equal to its IPC weight (censored observations had IPC weights of zero). IPC weights were calculated from the Kaplan–Meier estimate for the censoring distribution as recommended in Hothorn et al. (2006). RF regression with log-transformed time as the outcome was used to grow each tree. Predicted values were tree-weighted averaged values of log-transformed survival time using weights as described in Step 4 of Section 3.1 in Hothorn et al. (2006).

To estimate prediction error, 100 independent bootstrap samples of each data set were used. Each method was fit on the bootstrap data and prediction error was estimated using the corresponding OOB data. For RSF, this method of estimation differs from the OOB method discussed in Section 5. Although OOB error estimation could have been employed, not doing so by-passes a potential problem with the RF censored regression approach. Sometimes a case will receive a very large IPC weight and appear in all or nearly every bootstrap sample, thus precluding accurate OOB prediction for that case [Hothorn et al. (2006)].

When computing the C-index, each method used a different predicted outcome. For RSF, ensemble mortality was used. For RF regression and Cox regression, predicted survival time and the Cox linear predictor were used, respectively.

Results from our experiment are shown in Figure 1. Our findings are summarized as follows:

1. In nearly all examples, RSF using logrank and logrankscore splitting had the lowest prediction error. Conservation of events splitting was also very good.
2. Interestingly, RSF using the logrankrandom splitting rule was good in all examples. Because this splitting rule is significantly faster, its performance suggests it might be the preferred method in settings where computational speed is essential.
3. Prediction errors from breast10, breast50 and breast100 demonstrated all forest methods were stable in the presence of noise variables. Cox regression, in contrast, became progressively worse as the number of noise variables increased.
4. Performance of RF regression depended strongly on the censoring rate. For the transplant and veteran data, where almost all cases were deaths, performance was good. For the prostate and esophagus data, where censoring was higher, performance was poor.

**7. Variable importance (VIMP).** Variables can be selected by filtering on the basis of their variable importance (VIMP). To calculate VIMP for a variable $x$, drop OOB cases down their in-bag survival tree. Whenever a split for $x$ is encountered, assign a daughter node randomly. The CHF from each such tree is calculated and averaged. The VIMP for $x$ is the prediction error for the original ensemble subtracted from the prediction error for the new ensemble obtained using randomizing $x$ assignments [Breiman (2001), Ishwaran (2007)].

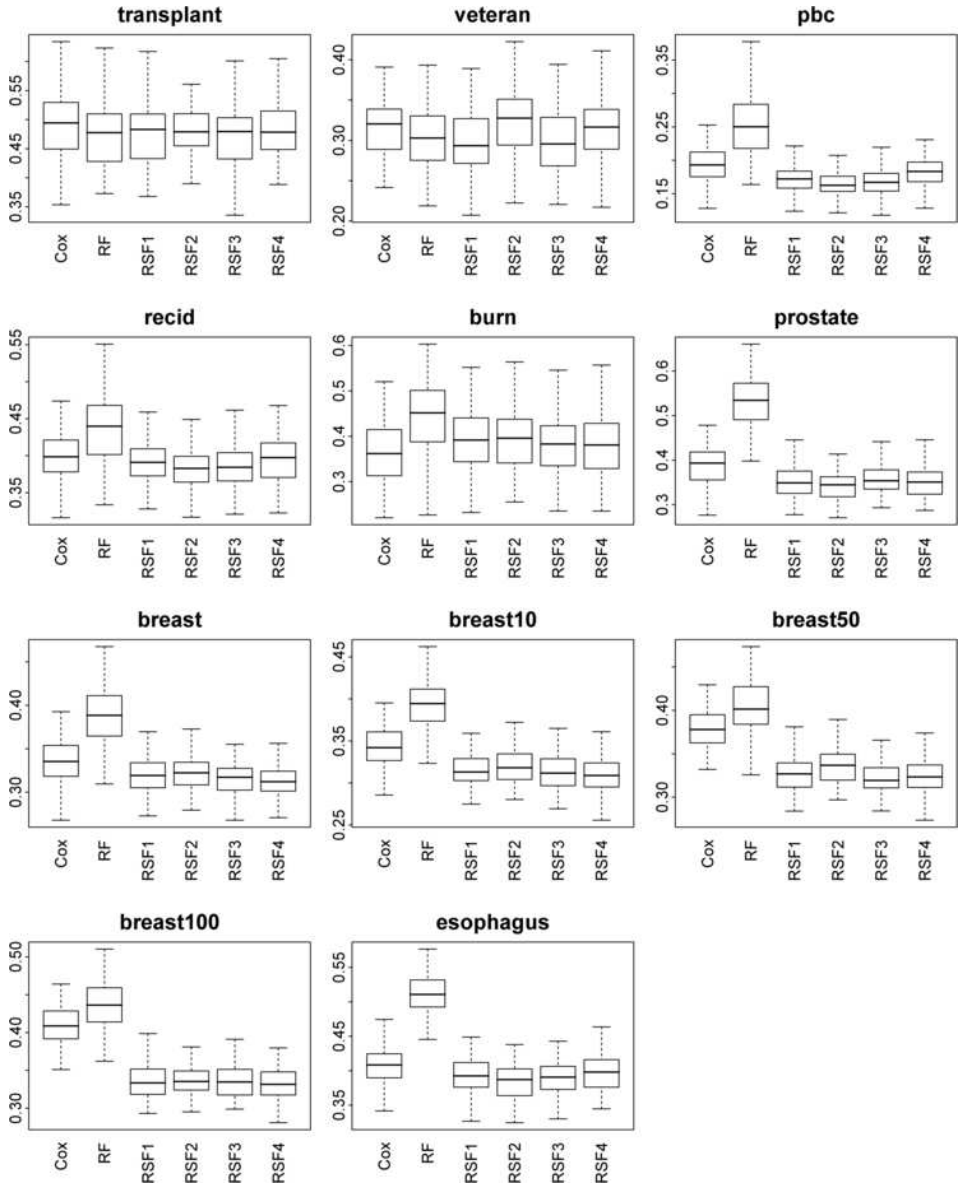

FIG. 1. *Boxplots of estimated prediction error (calculated using by C-index of Section 5) from 100 independent bootstrap replicates. Prediction error estimated on out-of-bag data. Dots in boxplots indicate mean values; horizontal lines are medians. Methods compared were as follows: Cox (Cox-regression); RF (RF for censored data [Hothorn et al. (2006)]; RSF1 through RSF4 (RSF using log-rank, conservation-of-events, log-rank score and random log-rank splitting). All forest analyses comprised 1000 trees. Datasets are indicated above each boxplot in bold.*

Large importance values indicate variables with predictive ability, whereas zero or negative values identify nonpredictive variables to be filtered. Under the C-index, one can interpret VIMP in terms of misclassification. Recall the C-index estimates the probability of correctly classifying two cases. Thus, VIMP for $x$ measures the increase (or drop!) in misclassification error on the test data if $x$ were not available.

One should be careful interpreting VIMP. Although tempting, it is incorrect to think VIMP estimates change in prediction error for a forest grown with and without a variable. For example, if two variables are highly correlated and both predictive, each can have large VIMP values. Removing one variable and regrowing the forest may affect the VIMP for the other variable (its value might get larger), but prediction error will likely remain unchanged. VIMP measures the change in prediction error on a fresh test case if $x$ were not available, given that the original forest was grown using $x$. Although, in practice, this often equals change in prediction error for a forest grown with and without $x$, conceptually the two quantities are different.

To examine the empirical properties of VIMP, we re-analyzed the breast cancer datasets used in Section 6. For each dataset, a forest of 1000 trees was grown using log-rank splitting. VIMP for each variable in each dataset was recorded. The analysis was repeated 100 times independently and VIMP averaged over the runs (Table 1).

Table 1 shows VIMP to be reasonably stable across datasets. In some instances, VIMP increases or decreases slightly for a variable as number of noise variables increases, but the ranking and relative size of VIMP is reasonably consistent. Also apparent from the table is the ability of VIMP to identify noise variables. The average absolute VIMP for noise variables in breast10, breast50 and breast100 is 0.001, and number of noise variables with VIMP exceeding 0.002 is 0.21 on average for breast100. This latter observation can be used as a means for thresholding variables, because any variable with a VIMP exceeding 0.002 is unlikely to be noise. Hence, we can conclude positive nodes, age, and progesterone are highly predictive, and hormone therapy and estrogen are moderately predictive. However, menopause, tumor size and tumor grade are unlikely to be predictive.

**8. Missing data.** One of the earliest algorithms for treating missing data in CART was based on the idea of a surrogate split [Chapter 5.3, Breiman et al. (1984)]. If $s$ is the best split for a variable $x$, the surrogate split for $s$ is the split $s^*$ using some other variable $x^*$ such that $s^*$ and $s$ are closest to one another in terms of predictive association [Breiman et al. (1984)]. To assign a case having a missing value for the variable used to split a node, the CART algorithm uses the best surrogate split among those variables not missing for the case. Surrogate splitting ensures every case can be classified, whether the case has missing values or not.

TABLE 1
*Variable importance (VIMP) for breast cancer datasets. Last two rows are mean of absolute VIMP for noise variables and number of noise variables with VIMP exceeding 0.002 (applies to breast10, breast50 and breast100). All reported values averaged over 100 independent runs. Each run based on 1000 trees under log-rank splitting*

|                  | breast | breast10 | breast50 | breast100 |
|------------------|--------|----------|----------|-----------|
| Hormone therapy  | 0.005  | 0.003    | 0.002    | 0.001     |
| Age              | 0.021  | 0.018    | 0.016    | 0.017     |
| Menopause        | −0.002 | 0.000    | 0.000    | 0.000     |
| Tumor size       | −0.002 | −0.002   | 0.000    | 0.001     |
| Tumor grade      | 0.003  | 0.001    | 0.001    | 0.001     |
| Positive nodes   | 0.037  | 0.040    | 0.041    | 0.042     |
| Progesterone     | 0.009  | 0.015    | 0.018    | 0.019     |
| Estrogen         | 0.003  | 0.003    | 0.004    | 0.005     |
| Noise [mean]     | —      | 0.002    | 0.001    | 0.001     |
| Noise [misclass] | —      | 0.090    | 0.190    | 0.210     |

Although surrogate splitting works well for trees, the method may not be well suited for forests. Speed is one issue. Finding a surrogate split is computationally intensive and may become infeasible when growing a large number of trees, especially for fully saturated trees used by forests. Further, surrogate splits may not even be meaningful in a forest paradigm. RF randomly selects variables when splitting a node and, as such, variables within a node may be uncorrelated, and a reasonable surrogate split may not exist. Another concern is that surrogate splitting alters the interpretation of a variable, which affects measures such as VIMP.

For these reasons, a different strategy is required for RF. The current method for imputing data uses a proximity approach [Breiman (2003), Liaw and Wiener (2002)]. This works as follows. First, the data are roughly imputed: missing values for continuous variables are replaced with the median of nonmissing values, or if the variable is categorical, data are imputed using the most frequent occurring nonmissing value. The roughly imputed data is analyzed using RF. Missing data are imputed using the resulting proximity matrix, an $n \times n$ symmetric matrix whose $(i,j)$ entry records the frequency that case $i$ and $j$ occur within the same terminal node. For continuous variables, imputed data are the proximity weighted average of the nonmissing data. For integer variables, imputed data are the integer value having the largest average proximity over nonmissing data. The updated data are then used as an input in RF, and the cycle is repeated. Typically, only a few iterations are needed to achieve a stable solution [Breiman (2003)].

An obvious advantage of such an approach is that it works off-the-shelf without special modification of the RF algorithm. Also, an important prop-erty of RF is its ability to cluster the data. Imputation via proximity exploits this. However, although the method is reasonable, there are some disadvantages:

1. OOB estimates for prediction error are biased, generally on the order of 10–20% [Breiman (2003)]. Further, because prediction error is biased, so are other measures based on it, such as VIMP.
2. The forest cannot be used to predict on test data with missing values.

8.1. *A new missing data algorithm: adaptive tree imputation.* To address these issues, we introduce a new missing data algorithm for forests. The idea is to adaptively impute missing data as a tree is grown. Imputation works by drawing randomly from the set of nonmissing in-bag data within the working node. Because only in-bag data are used for imputing, OOB data are not touched, and the OOB prediction error is not optimistically biased. The key steps involved are sketched below. For simplicity, we initially focus on the case where only $x$ variables have missing data:

1. For each node $h$, impute missing data prior to splitting. Let $\mathbf{X}^*_{k,h}$ be the set of nonmissing values for the $k$th coordinate of those **x**-variables in-bag in $h$. Let $\mathbb{P}^*_{k,h}$ be the empirical distribution function for $\mathbf{X}^*_{k,h}$. For each in-bag case in $h$ with missing value for the $k$th coordinate, impute by drawing a random value from $\mathbb{P}^*_{k,h}$. Repeat for each $k$. Splitting proceeds as usual once the data are imputed. Note carefully that only in-bag data is used as the basis for imputation and splitting.
2. The OOB data plays a passive role. It is imputed by drawing from $\mathbb{P}^*_{k,h}$.
3. Daughter nodes contain no missing data because the parent node is imputed prior to splitting. Reset the imputed data in the daughters to missing. Proceed as in Step 1 above, continuing until the tree can no longer be split.

The final summary imputed value for a missing case uses in-bag imputed values from the case's terminal nodes across the forest. If a case has a missing value for a continuous variable, the final summary value is the average of its imputed in-bag values. If a case has a missing value for an integer variable, the summary value is its in-bag imputed value occurring most frequently (in case of ties, a random tie-breaking rule is used).

8.2. *Imputation for test data.* The missing data algorithm can be used for test data. The test data are dropped down each tree of the forest and missing values are imputed dynamically as in Step 1 using $\mathbb{P}^*_{k,h}$. Once terminal node membership across the forest is established, missing data are summary imputed. If a test case has missing values for a continuous variable, the final summary value is its average imputed value. If a test case has missing values for an integer variable, the summary value is its imputed value occurring most frequently.

8.3. *Imputation for missing outcomes.* Imputation for missing values of censoring indicators and survival times proceeds in the same manner as for $x$ variables. Within the working node $h$, missing outcome data are drawn randomly from the empirical distribution function of the nonmissing in-bag outcome data. Once missing outcome data and missing values for $x$ variables are imputed, splitting proceeds using imputed in-bag data.

Summary imputation for outcomes is similar to that for $x$ variables. If survival time for a case is missing, the summary value is the average of its imputed in-bag values. If the censoring indicator for a case is missing, the summary value is its in-bag imputed censoring indicator occurring most frequently. OOB prediction error is calculated using imputed outcomes. If a case has a missing outcome, its OOB summary imputed value is used when calculating the prediction error.

Test data with missing outcomes are treated as follows. Test data are dropped down each tree and missing data are imputed within a node by drawing a random value from the empirical distribution function of the nonmissing in-bag training outcome data. Final summary imputation uses imputed values of test cases from terminal nodes.

8.4. *Iterating the missing data algorithm.* With increasing amounts of missing data, accuracy of imputed values may degrade. In such cases, accuracy can be improved substantially by iterating the missing data algorithm. Iterative imputation works as follows. After the initial cycle of growing the forest, missing data are imputed using OOB summary values. A new forest is then grown using the imputed data. For each case originally missing a value, draw a random value from the nonmissing in-bag cases within its terminal node. A value is drawn for each tree in the new forest. Impute cases with missing data using full (in-bag plus OOB data) summary imputation. Use the reimputed data and grow another forest. Repeat iteratively.

8.5. *Empirical performance.* To illustrate the missing data algorithm, we use the pbc data considered in Section 6. The data is from a clinical trial involving primary biliary cirrhosis (an autoimmune disease of the liver) and comprise 312 individuals who participated in a randomized trial to study the effectiveness of the drug D-penicillamine. The dataset contains 17 variables, as well as censoring information and time until death for each individual [Fleming and Harrington (1991)].

A preliminary analysis of the data using RSF (1000 trees under log-rank splitting) identified several variables that have large VIMP values. VIMP for the top predictive variable, serum bilirubin, was especially large, being almost twice the size of the second most predictive variable, age. This finding is consistent with biology. In normal patients, a small amount of bilirubin, a waste produced by the breakdown of old red blood cells and hemoglobin, circulates in the blood, but elevated amounts are found in patients with liver disease. Serum bilirubin level is considered an accurate test of liver function. Therefore, it is not surprising that this variable is predictive of disease.

Given the importance of serum bilirubin, we were curious to see how well we could impute this value and what effects imputation might have on prediction error. Because the variable had no missing data (in fact, few data were missing), we randomly assigned missing values. In doing so, we first put aside a 20% subset of the data for testing. Over the remaining data, missing values were randomly assigned to variables. The missing data algorithm was applied to these data, and the root-mean-square error (RMSE) between the imputed and true value for serum bilirubin was calculated. The OOB error rate from the training data and the test set error were also calculated.

For comparison, we iterated the missing data algorithm. We also used proximity imputation [Breiman (2003), Liaw and Wiener (2002)]. A five-iteration cycle was used in both cases. Results for all three methods are reported in Table 2. Values given are averaged over 100 random datasets generated under differing amounts of missingness.

Several trends can be observed from Table 2:

1. Imputed values were accurate and OOB error matched test set error for the missing data algorithm under moderate amounts of missing data (5–10%). As missing data increased (25–50%), OOB error overestimated test set error and RMSE performance degraded.
2. The iterative missing data algorithm was consistently good. RMSE was consistently low and OOB error closely matched test set error in all settings.

TABLE 2
*Imputation for pbc data under differing amounts of missing values. Reported for each method are the following:* (i) *root-mean-square error (RMSE) between imputed and actual values for the covariate serum bilirubin (for comparison note that serum bilirubin had standard deviation* 4.41); (ii) *OOB prediction;* (iii) *test set prediction error using a 20% hold out test set* (*TEST*). *Values are averaged over 100 random datasets. Forests comprised 1000 trees under log-rank splitting*

| %<br>missing<br>data | Missing data<br>algorithm | | | Missing data<br>algorithm<br>(5 iterations) | | | Proximity<br>imputation<br>(5 iterations) | | |
|---|---|---|---|---|---|---|---|---|---|
| | RMSE | OOB | TEST | RMSE | OOB | TEST | RMSE | OOB | TEST |
| 5 | 3.181 | 0.170 | 0.174 | 3.033 | 0.165 | 0.175 | 3.068 | 0.171 | 0.174 |
| 10 | 3.537 | 0.173 | 0.168 | 3.323 | 0.163 | 0.169 | 3.381 | 0.173 | 0.166 |
| 25 | 3.766 | 0.185 | 0.165 | 3.357 | 0.159 | 0.169 | 3.420 | 0.173 | 0.165 |
| 50 | 3.912 | 0.204 | 0.155 | 3.416 | 0.156 | 0.161 | 3.464 | 0.164 | 0.157 |

3. Proximity imputation was also consistently good. Imputed values were accurate, and OOB error rates matched test set errors in all settings. We did not notice bias in OOB error rates reported elsewhere [Breiman (2003)].

These findings suggest the missing data algorithm can be used confidently with low to moderate missing data. Its advantages include the ability to predict on test data with missing values and that it can be used when outcomes are missing. With increasing amounts of missing data, the algorithm can still be used, but should be iterated a few times to improve accuracy of imputed values. Interestingly, iterating the algorithm did not bias OOB error rates in our simulations. Informal experiments with other data showed this pattern to be consistent, suggesting that the use of OOB summary imputation on the first iteration and full summary imputation on further iterations (which mixes in-bag and OOB data) is helping to mitigate bias. We believe this same effect is also at play for proximity imputation. Proximity summary imputation uses both in-bag and OOB data, and this must be helping to reduce bias. The results for proximity imputation were better than expected and suggest it could also be used for missing data.

**9. Body mass index and long-term survival among patients undergoing coronary artery bypass grafting (CABG) surgery.** Over the past twenty years, public health investigators have documented a dramatic increase in the proportion of the US population that is overweight or obese [Mokdad et al. (2003)]. A common method of assessing obesity is the calculation of body mass index, which is weight (in kilograms) divided by height (in meters) squared ($kg/m^2$). Body mass index values of $< 18.5$, 18.5–25, 25–30, 30–35 and $> 35$ correspond to underweight, normal weight, overweight, grade I obesity and grade II obesity [Flegal (2005, 2007)].

The association of body mass with mortality is complex. People who are underweight are at increased risk [Adams et al. (2006), Flegal et al. (2005)], but it is unclear whether this is due to reverse causality from chronic conditions such as renal dysfunction or undiagnosed cancer. Smoking decreases body mass but increases death risk from coronary artery disease, lung cancer, obstructive lung disease, and other diseases. High body mass index predicts increased risk, but it is not clear whether risk starts to increase when subjects become simply overweight or frankly obese [Adams et al. (2006), Flegal et al. (2005)].

Obesity increases the risk of developing and dying from coronary artery disease [Flegal et al. (2007)], at least in part by predisposing people to diabetes and hypertension. Although it is widely accepted that obesity increases the risk of developing disease, there is controversy over the prognostic implications of obesity among patients with established disease; some suggest there may even be a paradoxical protective effect of obesity [Urtesky et al. (2007)].

In order to explore these complex patterns relating obesity to mortality in patients with stable, documented, severe coronary artery disease, we focus on patients who underwent primary isolated CABG surgery. CABG was developed in the late 1960s for treating refractory angina and has been shown to improve life expectancy in patients with severe obstructive coronary artery disease. Early randomized trials demonstrated that compared with medical therapy, CABG resulted in better survival [Yusuf, Zucker and Peduzzi (1994)]. Since its inception, the procedure has evolved substantially, with greater use of arterial revascularization and incorporation of minimally invasive techniques [Puskas et al. (2004)]. A number of models have been constructed to predict long-term survival after CABG, but none has included body mass index as a predictor [Eagle et al. (2001)].

The data analyzed comprised 15,586 patients accrued between the years 1990 and 2003 at Cleveland Clinic. The outcome used for the analysis was all cause mortality. Mean follow-up time was $6.5 \pm 3.8$ years, the median was 6.4 years. In total there were 36 variables analyzed, including: body mass index; age in years; creatinine clearance, a continuous measure of renal function; smoking history; and number of internal thoracic artery grafts used for CABG. These latter five variables were found to be among the most predictive using a RSF analysis.

Five-year predicted survival (estimated from the ensemble) appears in Figure 2 plotted against body mass index. Survival has been conditioned on smoking history and creatinine clearance level. Solid lines appearing in the figure are lowess smoothed estimates of survival stratified by number of internal thoracic artery grafts used.

One can see a distinctive "hockey stick" pattern in Figure 2. At low body mass index, survival is low, after which survival increases with increasing body mass until reaching an inflection point of roughly 25 kg/m$^2$, where it then begins to decrease. Interestingly, this pattern is highly dependent on creatinine clearance levels. For creatinine clearance values larger than 90 ml/min, signifying healthy renal function, the hockey stick pattern is much straighter (coplots on the extreme right-hand side).

These results add strength to the hypothesis that there is a reverse causation effect for underweight individuals. The poor survival seen in patients with low body mass index could be construed as an effect of being underweight, but this may be incorrect. As seen, the association of low body mass with survival is related to renal function, and this effect dissipates as renal function improves. Thus, poor survival in patients with low body mass index may not be due to being underweight, but rather due to the systemic effects of renal disease.

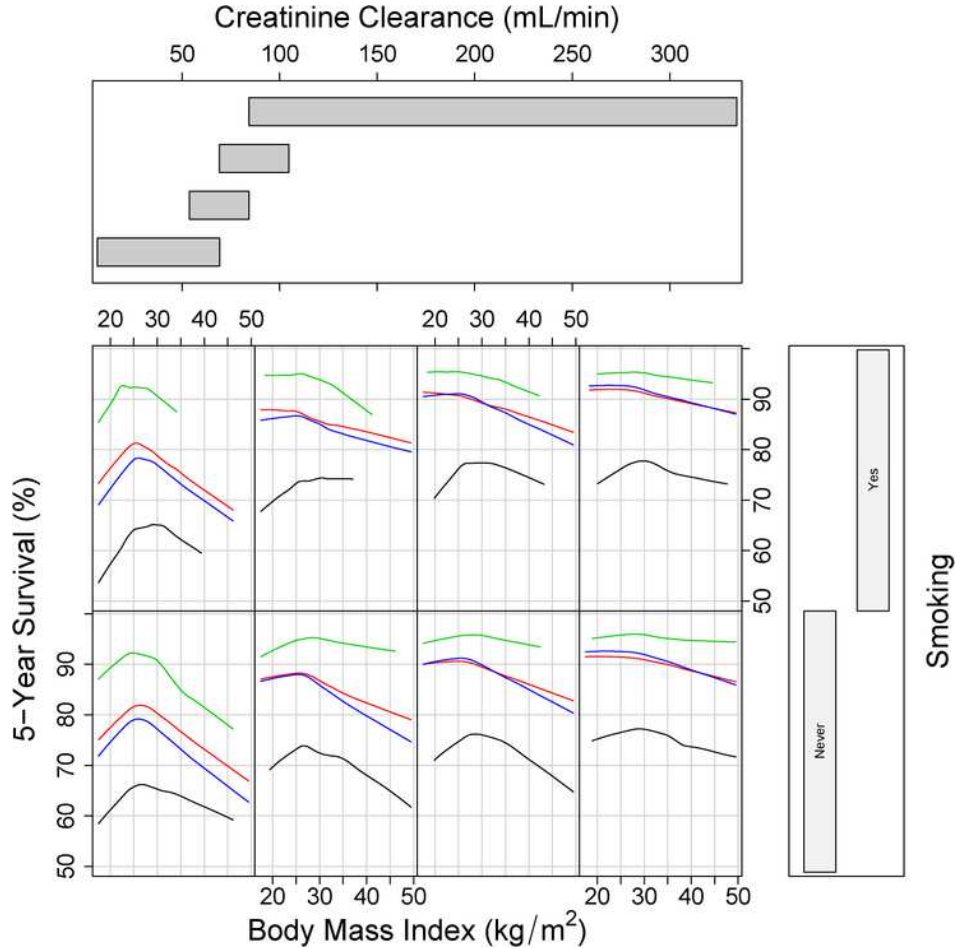

Fig. 2. *Predicted five-year survival versus body mass index ($kg/m^2$), conditioned on creatinine clearance (ml/min) and smoking history. Curves are smoothed using lowess and stratified by number of internal thoracic artery grafts (black = 0; red = 1; green = 2; blue = combined). Analysis of CABG data based on RSF using 1000 trees with log-rank splitting.*

The effect of smoking on survival is also interesting. Smoking decreases body mass, but at the same time leads to increased risk and poorer survival. This effect can be seen in Figure 2. For patients with no internal thoracic artery grafts (black curves) and normal renal function (creatinine clearance > 90 ml/min), smoking leads to a more pronounced hockey stick pattern (top right coplot versus bottom right coplot). Note that for the same range of creatinine clearance, there is no effect when patients have at least one internal thoracic artery graft (green and red curves). These patients have additional coronary artery protection, and the protection that this confers masks the lesser effect of smoking. The effect of smoking on survival is also less apparent in patients with abnormal creatinine clearance levels. Again, this is because the effect is being masked, in this case by renal dysfunction.

**10. Discussion.** In this paper we have introduced RSF, a new extension of Breiman's forests method [Breiman (2001)], to right-censored survival data. A random survival forest consists of random survival trees. Using independent bootstrap samples, each tree is grown by randomly selecting a subset of variables at each node and then splitting the node using a survival criterion involving survival time and censoring status information. The tree is considered fully grown when each terminal node has no fewer than $d_0 > 0$ unique deaths. The estimated CHF for a case is the Nelson–Aalen estimator for the case's terminal node. The ensemble is the average of these CHFs. Because trees are grown from in-bag data, an OOB ensemble can be calculated by dropping OOB cases down their in-bag survival trees and averaging. The predicted value for a case using the OOB ensemble does not use survival information for that case, and, therefore, it can be used for nearly unbiased estimation of prediction error. From this, other useful measures can be derived, such as VIMP values for filtering and selecting variables.

RSF incorporates many of the useful ideas promoted by Breiman (2001). At the same time, we have proposed new ways to extend the methodology. A novel missing data algorithm was introduced that can be used for both training and testing data and that provides nearly unbiased estimates for error rates even with large amounts of missing data.

A large experiment was used to assess prediction accuracy of RSF. Over a wide range of real as well as simulated datasets, we found RSF to be consistently better than, or at least as good as, competing methods. Since the introduction of RF to the machine learning community, there has been a tremendous effort to document its empirical performance. Our results confirm what has generally been found: RF produces highly accurate ensemble predictors.

We have also illustrated the ease with which RSF can be applied to real data settings to uncover highly complex interrelationships between variables. Our case study involving coronary artery disease found important relationships among renal function, body mass index, and long-term survival that help explain much of the confusion reported in the literature on this controversial topic. Complex relationships like this are found with ease using tools such as VIMP in combination with the highly adaptive nature of forests. In contrast, conventional methods are much less automatic and require considerable subjective input from the user in data settings where variables are highly interrelated.

**Acknowledgments.** The authors thank Michael Stein, the Associate Editor, and two anonymous referees who all helped to substantially improve the paper.

## REFERENCES

ADAMS, K. F., SCHATZKIN, A., HARRIS, T. B. ET AL. (2006). Overweight, obesity, and mortality in a large prospective cohort of persons 50 to 71 years old. *N. Engl. J. Med.* **355** 763–778.

BREIMAN, L. (1996). Bagging predictors. *Machine Learning* **26** 123–140.

BREIMAN, L. (2001). Random forests. *Machine Learning* **45** 5–32.

BREIMAN, L. (2002). Software for the masses. Slides presented at the Wald Lectures, Meeting of the Institute of Mathematical Statistics, Banff, Canada. Available at http://www.stat.berkeley.edu/users/breiman.

BREIMAN, L. (2003). Manual—setting up, using and understanding random forests V4.0. Available at ftp://ftp.stat.berkeley.edu/pub/users/breiman/Using_random_forests_v4.0.pdf.

BREIMAN, L., FRIEDMAN, J. H., OLSHEN, R. A. and STONE, C. J. (1984). *Classification and Regression Trees*. Wadsworth, Belmont, California. MR0726392

CORTES, C. and VAPNIK, V. N. (1995). Support-vector networks. *Machine Learning* **20** 273–297.

EAGLE, K. A., GUYTON, R. A., DAVIDOFF, R. ET AL. (2004). ACC/AHA 2004 guideline update for coronary artery bypass graft surgery: Summary article. A report of the American College of Cardiology/American Heart Association Task Force on Practice Guidelines (Committee to Update the 1999 Guidelines for Coronary Artery Bypass Graft Surgery). *J. Am. Coll. Cardiol.* **44** 213–310.

FLEGAL, K. M., GRAUBARD, B. I., WILLIAMSON, D. F. and GAIL, M. H. (2005). Excess deaths associated with underweight, overweight and obesity. *J. Amer. Med. Assoc.* **293** 1861–1867.

FLEGAL, K. M., GRAUBARD, B. I., WILLIAMSON, D. F. and GAIL, M. H. (2007). Cause-specific excess deaths associated with underweight, overweight and obesity. *J. Amer. Med. Assoc.* **298** 2028–2037.

FONTAINE, K. R., REDDEN, D. T., WANG, C., WESTFALL, A. O. and ALLISON, D. B. (2003). Years of life lost due to obesity. *J. Amer. Med. Assoc.* **289** 187–193.

FLEMING, T. and HARRINGTON, D. (1991). *Counting Processes and Survival Analysis*. Wiley, New York. MR1100924

HARRELL, F., CALIFF, R., PRYOR, D., LEE, K. and ROSATI, R. (1982). Evaluating the yield of medical tests. *J. Amer. Med. Assoc.* **247** 2543–2546.

HEAGERTY, P. J. and ZHENG, Y. (2005). Survival model predictive accurracy and ROC curves. *Biometrics* **61** 92–105. MR2135849

HOTHORN, T. and LAUSEN, B. (2003). On the exact distribution of maximally selected rank statistics. *Comput. Statist. Data Anal.* **43** 121–137. MR1985332

HOTHORN, T., BUHLMANN, P., DUDOIT, S., MOLINARO, A. and VAN DER LAAN, M. J. (2006). Survival ensembles. *Biostat.* **7** 355–373.

ISHWARAN, H. (2007). Variable importance in binary regression trees and forests. *Electron. J. Statist.* **1** 519–537. MR2357716

ISHWARAN, H., BLACKSTONE, E. H., POTHIER, C. and LAUER, M. S. (2004). Relative risk forests for exercise heart rate recovery as a predictor of mortality. *J. Amer. Statist. Assoc.* **99** 591–600. MR2086385

Ishwaran, H. and Kogalur, U. B. (2007). Random survival forests for R. *Rnews* **7** 25–31.

Ishwaran, H. and Kogalur, U. B. (2008). RandomSurvivalForest 3.2.2. R package. Available at http://cran.r-project.org.

Ishwaran, H., Blackstone, E. H., Apperson, C. A. and Rice, T. W. A novel data-driven approach to stage grouping of esophageal cancer. Cleveland Clinic technical report.

Kalbfleisch, J. and Prentice, R. (1980). *The Statistical Analysis of Failure Time Data*. Wiley, New York. MR0570114

Kattan, M. (2003). Comparison of Cox regression with other methods for determining prediction models and nomograms. *J. Urol.* **170** S6–S10.

LeBlanc, M. and Crowley, J. (1992). Relative risk trees for censored survival data. *Biometrics* **48** 411–425.

LeBlanc, M. and Crowley, J. (1993). Survival trees by goodness of split. *J. Amer. Statist. Assoc.* **88** 457–467. MR1224370

Liaw, A. and Wiener, M. (2002). Classification and regression by randomForest. *Rnews* **2/3** 18–22.

Liaw, A. and Wiener, M. (2007). RandomForest 4.5-18. R package. Available at http://cran.r-project.org.

Molinaro, A. M., Dudoit, S. and van der Laan, M. J. (2004). Tree-based multivariate regression and density estimation with right-censored data. *J. Multivariate Anal.* **90** 154–177. MR2064940

Mokdad, A. H, Ford, E. S., Bowman, B. A. et al. (2003). Prevalence of obesity, diabetes and obesity-related health risk factors. *J. Amer. Med. Assoc.* **289** 76–79.

Naftel, D., Blackstone, E. H. and Turner, M. (1985). Conservation of events. Unpublished notes.

Olshansky, S. J., Passaro, D. J., Hershow, R. C. et al. (2005). A potential decline in life expectancy in the United States in the 21st century. *N. Engl. J. Med.* **352** 1138–1145.

Puskas, J. D., Williams, W. H., Mahoney, E. M. et al. (2004). Off-pump vs conventional coronary artery bypass grafting: Early and 1-year graft patency, cost and quality-of-life outcomes: A randomized trial. *J. Amer. Med. Assoc.* **291** 1841–1849.

Rossi, P. H., Berk, R. A. and Lenihan, K. J. (1980). *Money, Work and Crime: Some Experimental Results*. Academic Press, New York.

Schapire, R., Freund, Y., Bartlett, P. and Lee, W. (1998). Boosting the margin: A new explanation for the effectiveness of voting methods. *Ann. Statist.* **26** 1651–1686. MR1673273

Segal, M. R. (1988). Regression trees for censored data. *Biometrics* **44** 35–47.

Uretsky, S., Messerli, F. H., Bangalore, S. et al. (2007). Obesity paradox in patients with hypertension and coronary artery disease. *Am. J. Med.* **120** 863–870.

Yusuf, S., Zucker, D., Peduzzi, P. et al. (1994). Effect of coronary artery bypass graft surgery on survival: Overview of 10-year results from randomised trials by the Coronary Artery Bypass Graft Surgery Trialists Collaboration. *Lancet* **344** 563–570.

H. Ishwaran  
Department of Quantitative  
Health Sciences  
Cleveland Clinic  
9500 Euclid Avenue  
Cleveland, Ohio 44195  
USA  
E-mail: hemant.ishwaran@gmail.com

U. B. Kogalur  
Department of Statistics  
Columbia University  
1255 Amsterdam Avenue  
New York, New York 10027  
USA  
E-mail: ubk2101@columbia.edu

E. H. Blackstone
Department of Thoracic
  and Cardiovascular Surgery
Cleveland Clinic
9500 Euclid Avenue
Cleveland, Ohio 44195
USA
E-mail: blackse@ccf.org

M. S. Lauer
Division of Prevention
  and Population Science
National Heart, Lung, and Blood Institute
Rockledge Center II
6701 Rockledge Drive
Besthesda, Maryland 20892
USA
E-mail: lauerm@nhlbi.nih.gov